\documentclass[twocolumn,showpacs,prb,amsmath,amssymb]{revtex4-2}

\usepackage{graphicx}
\usepackage{dcolumn}
\usepackage{bm}
\usepackage{color}

\newcommand{\subm}[1]{_{\mathrm {#1}}}

\begin{document}

\preprint{APS/123-QED}
\title{Spectroscopic Evidence for Electron-Boson Coupling in Half-metallic CrO$_2$}

\author{Daiki~Ootsuki$^1$}
\email{ootsuki.daiki@okayama-u.ac.jp}
\author{Hirokazu~Fujiwara$^2$}
\author{Noriyuki~Kataoka$^3$}
\author{Kensei~Terashima$^4$}
\author{Miho~Kitamura$^{5},^{\dagger}$}
\author{Koji~Horiba$^5$}
\thanks{Present affiliation: Institute for Advanced Synchrotron Light Source, National Institutes for Quantum and Radiological Science and Technology, 6-6-11 Aoba, Sendai, Miyagi, 980-8579, Japan.}
\author{Hiroshi~Kumigashira$^6$}
\author{Shiv~Kumar$^7$}
\author{Shin-ichiro~Ideta$^7$}
\author{Kenya~Shimada$^{7,8,9}$}
\author{Yuji~Muraoka$^1$}
\author{Takayoshi~Yokoya$^1$}

\affiliation{$^1$Research Institute for Interdisciplinary Science, Okayama University, Okayama 700-8530, Japan} 
\affiliation{$^2$Department of Advanced Materials Science, Graduate School of Frontier Sciences, and Material Innovation Research Center (MIRC), The University of Tokyo, Chiba 277-8561, Japan}
\affiliation{$^3$Graduate School of Natural Science and Technology, Okayama University, Okayama 700-8530, Japan}
\affiliation{$^4$Research Center for Materials Nanoarchitectonics (MANA), National Institute for Materials Science, 1-2-1 Sengen, Tsukuba, Ibaraki 305-0047, Japan}
\affiliation{$^5$Institute of Materials Structure Science, High Energy Accelerator Research Organization (KEK), Tsukuba, Ibaraki 305-0801, Japan}
\affiliation{$^6$Institute of Multidisciplinary Research for Advanced Materials (IMRAM), Tohoku University, Sendai 980-8577, Japan}
\affiliation{$^7$Research Institute for Synchrotron Radiation Science (HiSOR), Hiroshima University, Higashi-hiroshima 739-0046, Japan}
\affiliation{$^8$International Institute for Sustainability with Knotted Chiral Meta Matter (WPI-SKCM$^2$), Hiroshima University, Higashi-Hiroshima, Hiroshima 739-8526, Japan}
\affiliation{$^9$Research Institute for Semiconductor Engineering (RISE), Hiroshima University, Higashi-Hiroshima, Hiroshima 739-8527, Japan}

\date{\today}
            
\begin{abstract}
We report quasiparticle properties of the half-metal ferromagnet CrO$_2$ by means of high-resolution angle-resolved photoemission spectroscopy (ARPES). 
We clearly observed the Fermi surface (FS) and band dispersion in good agreement with the previous reports. 
Moreover, the ARPES band dispersion reveals a distinct kink structure around 68 meV, providing the first spectroscopic evidence for the elementary excitations in CrO$_2$.  
The energy scale of this feature is comparable to the Debye temperature and the $A\subm{1g}$ phonon mode, suggesting the electron-phonon interaction. 
From the detailed analysis, we have extracted the self-energy and found two characteristic structures in the real part of the self-energy. 
Assuming the existence of the electron-magnon interaction as well as the electron-phonon interaction, we could reproduce the evaluated real and imaginary parts of the self-energy as well as ARPES intensity. 
Our findings reveal the renormalized quasiparticle (QP) dynamics in CrO$_2$ and provide valuable insights into the fundamental many-body interactions governing half-metallic ferromagnets. 
\end{abstract}

\maketitle

\section{INTRUDUCTION}
A half-metal is a ferromagnetic material that is metallic for majority spin states but insulating/semiconducting for minority spin states.
Over the past 50 years, the half metallic behaviors have been proposed in various compounds such as transition-metal oxides \cite{Goodenough1971, Schwarz1986, Korotin1998, Korotin1998, Ji2001, Park1998,Dedkov2002}, spinels \cite{Horikawa1982}, Heusler alloys \cite{Hanssen1990, Jourdanetal2014, Kono2020}, and zinc blende \cite{Akinaga2000}. 
Investigating half-metallic ferromagnets remains important, as their perfect spin polarization plays a key role in spintronic applications through tunnel magnetoresistance.

Chromium dioxide CrO$_2$ is a promising half-metal ferromagnet with a nearly 100 \% spin polarization at the Fermi level ($E\subm{F}$) \cite{Soulen1998, Anguelouch2001} that has been extensively researched as a next-generation spintronics material. 
The half-metallic nature of CrO$_2$ has been predicted from the pioneering theoretical work by Schwarz \cite{Schwarz1986}, and the large spin polarization up to 98.4 \% at low temperatures has been confirmed experimentally in the point contact Andreev reflection measurements \cite{Soulen1998, Anguelouch2001}. 
The half-metallic nature leads to characteristic physical properties such as the anomalous temperature dependence of electrical resistivity, Hall coefficient, and magnetoresistance \cite{Lewis1997,Barry1998, Suzuki1998, Watts2000, Anwar2013}. 
From the perspective of electronic structures, x-ray absorption spectroscopy, infrared spectroscopy, and photoemission spectroscopy have revealed the half-metallic states \cite{Huang2002, Huang2003, Chang2005, Dedkov2005, Singley1999, Tsujioka1997, Sperlich2013}. 
Recently, the observations of the three-dimensional half-metallic electronic structure using soft x-ray ARPES and Shubnikov-de Haas oscillation have been reported \cite{Bisti2017,Bheemarasetty2025}. 
In particular, the spin-resolved photoemission spectroscopies have provided strong evidence for its highly spin-polarized electronic states near the Fermi level and anomalous thermal spin depolarization \cite{Kamper1987,Dedkov2002_2,Fujiwara2015,Fujiwara2018}. 
The nearly perfect spin-polarized state of CrO$_2$ leads to the peculiar temperature dependence of the electrical resistivity $\rho = \rho_0 + A T^2 e^{-\Delta/T}$ at low temperatures \cite{Barry1998,Suzuki1998,Watts2000,Anwar2013}. 
This behavior originates from the restriction of spin-flip scattering of electrons by magnon excitation in the absence of the minority spin state. 
It is different from the Fermi-liquid picture $\rho = \rho_0 + A T^2$ due to the electron-electron correlation in the normal ferromagnets. 
On the other hand, the electron-phonon interaction in CrO$_2$ has been discussed in terms of the Bloch-Gr\"{u}neisen analysis of $\rho$ \cite{Lewis1997}. 
Thus, the electron-phonon interaction as well as the electron-magnon interaction would play important roles in the nearly perfect spin-polarized states. 
Nevertheless, the interactions between electrons and these elementary excitation modes, which play a crucial role in transport properties, remains unconfirmed from the viewpoint of direct electronic structure observation. 

In this context, we have performed a high-resolution ARPES study for CrO$_2$ in order to clarify the QP states related to the spin-polarized ground state. 
We have successfully observed the clear Fermi surfaces, which is consistent with the previous SX-ARPES study \cite{Bisti2017}. 
Moreover, a clear kink structure in the ARPES spectrum along the $\Gamma^\prime$-X$^\prime$ direction was observed, indicating the QP renormalization phenomena. 
The observed energy scale $68$ meV of the kink structure is comparable to the Debye temperature $\Theta\subm{D} = 750$ K  as well as the $A\subm{1g}$ phonon mode of $72$ meV \cite{Lewis1997, Iliev1999}. 
Moreover, we have estimated the self-energy from the ARPES band dispersion and simulated the self-energy and the ARPES spectrum, assuming the electron-phonon and electron-magnon interactions. 
These results suggest the existence of two different bosonic modes coupled to electrons and provide direct spectroscopic evidence for the QP renormalization in CrO$_2$,  highlighting the crucial role of many-body interactions in its spin-polarized electronic structure.

\section{EXPERIMENTAL SETUP}
The CrO$_2$ (100) epitaxial films on a rutile-type TiO$_2$ (100) substrate were grown by a closed-system chemical vapor deposition (CVD) method \cite{Iwai2010}. 
ARPES measurements were carried out at BL-2A of Photon Factory with a Scienta SES2002 electron analyzer\cite{Horiba2003} and BL-1 of HiSOR with a Scienta Omicron R4000 electron analyzer. 
The incident photon energy was set to $h\nu$ = 114 eV with the circularly polarized light and the linearly polarized light. 
The total energy resolution was $\sim$ 30 meV for $h\nu$ = 114 eV. 
The base pressure of the chamber was about $1.0\times10^{-10}$ Torr. 
The binding energy was calibrated by using the Fermi edge of the gold reference. 
The data were collected at $T$ = 17 K and 20 K.

\section{RESULTS AND DISCUSSION}
\begin{figure}
\includegraphics[width=8.5cm]{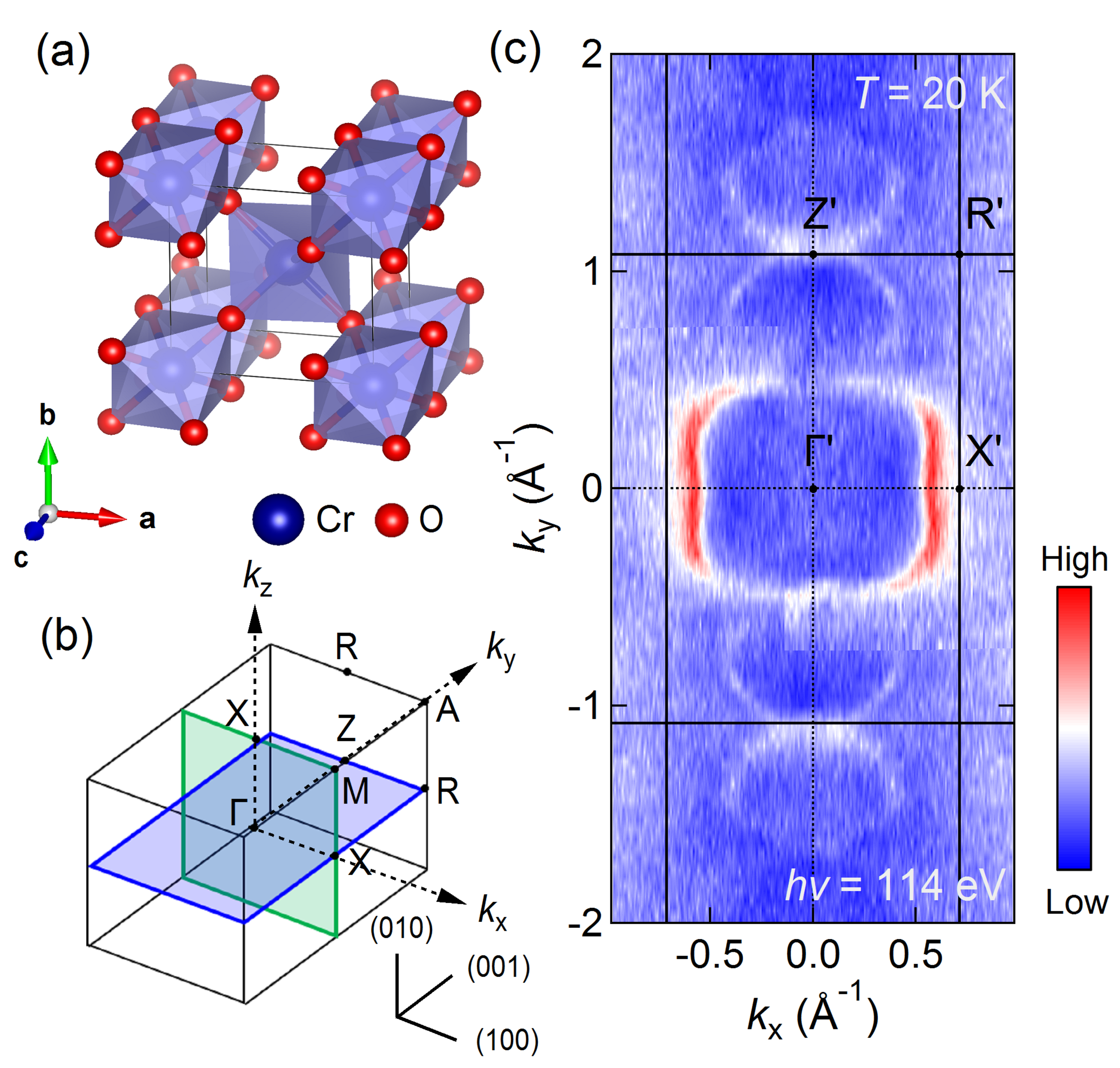}
\caption{(color online) (a) Crystal structure of CrO$_2$ visualized by VESTA \cite{vesta}. 
(b) Brillouin zone of CrO$_2$. 
(c) Symmetrized FS of CrO$_2$ at $T = 20$ K of $\Gamma^\prime$X$^\prime$R$^\prime$Z$^\prime$ plane, corresponding to the blue shaded area in (b). 
The data were collected at $h\nu = 114$ eV. 
 } 
\label{f1}
\end{figure}

\begin{figure}
\includegraphics[width=8.7cm]{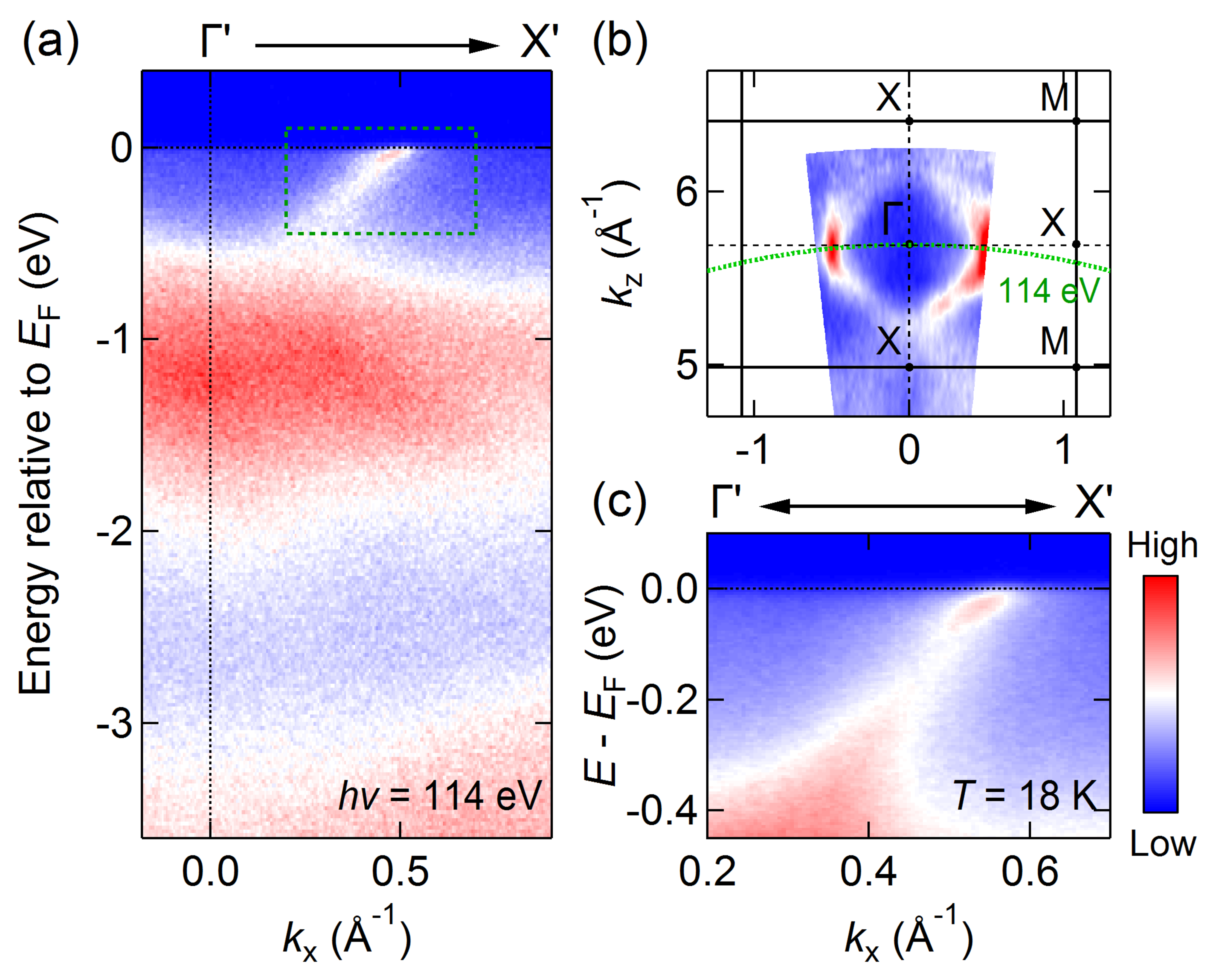}
\caption{(color online) (a) ARPES band dispersion along $\Gamma^\prime$-X$^\prime$ direction at $T$ = 20 K. 
(b) Photon-energy dependence of ARPES spectra in $\Gamma^\prime$X$^\prime$M$^\prime$ plane as indicated by the green shaded area in Fig. \ref{f1}(b). 
(c) Enlarged view of ARPES spectra near $E\subm{F}$ for the rectangular region (green dashed line) in (a). 
The ARPES spectra of (a) and (c) were obtained at $h\nu = 114$ eV, corresponding to the green dotted line in (b).  
} 
\label{f2}
\end{figure}

The rutile-type crystal structure of CrO$_2$ and its Brillouin zone (BZ) are displayed in Figs. \ref{f1}(a) and \ref{f1}(b). 
The magnetic easy axis is the (001) direction of (a). 
Figure \ref{f1}(c) shows the Fermi surface (FS) of CrO$_2$ at $T = 20$ K, approximately corresponding to the $\Gamma$XRZ plane in Fig. \ref{f1}(b). 
The data were collected at $h\nu = 114$ eV. 
Due to the relatively high energy resolution of vacuum-ultraviolet (VUV) light, the clear rectangular FS around $\Gamma^\prime$ was revealed. 
On the other hand, the figure-eight-shaped FS was also observed around $Z^\prime$. 
These results are consistent with the fully spin-polarized calculation and the FSs reported by the soft x-ray ARPES \cite{Bisti2017}. 
Generally, the probing depth of VUV light is shallow, and the large probing depth of soft x-ray is needed to obtain the ARPES data. 
However, this study demonstrates that clear FSs can also be observed even in the VUV region. 
This indicates that the ARPES image can be obtained with higher momentum resolution compared to the soft X-ray region.

Figure \ref{f2}(a) shows the ARPES band dispersion along $\Gamma^\prime$-X$^\prime$ direction and exhibits the electron band consisting of the rectangular FS in Fig. \ref{f1}(c). 
To confirm that the observed FS (electron band) in Fig. \ref{f1}(c) (Fig. \ref{f2}(a))  corresponds to the $\Gamma^\prime$X$^\prime$R$^\prime$Z$^\prime$ plane ($\Gamma$-X direction), we have checked the photon-energy dependence of the ARPES spectra along $\Gamma^\prime$-X$^\prime$ direction as shown in Fig. \ref{f2}(b). 
The rounded rectangular FS can be seen and agrees well with the previous SX-ARPES study \cite{Bisti2017}. 
Here, it turns out that the photon energy $h\nu = 114$ eV almost matches the $\Gamma$ point. 
The observed three-dimensional FS indicates that our VUV-ARPES results reflect the bulk electronic state of CrO$_2$. 
To carefully focus on the QP state near $E\subm{F}$, we have zoomed the ARPES band dispersion as shown in Fig. \ref{f2}(c). 
The intensity of the QP state very close to $E\subm{F}$ was strengthened, suggesting a significant many-body correlation.

\begin{figure*}
\includegraphics[width=17cm]{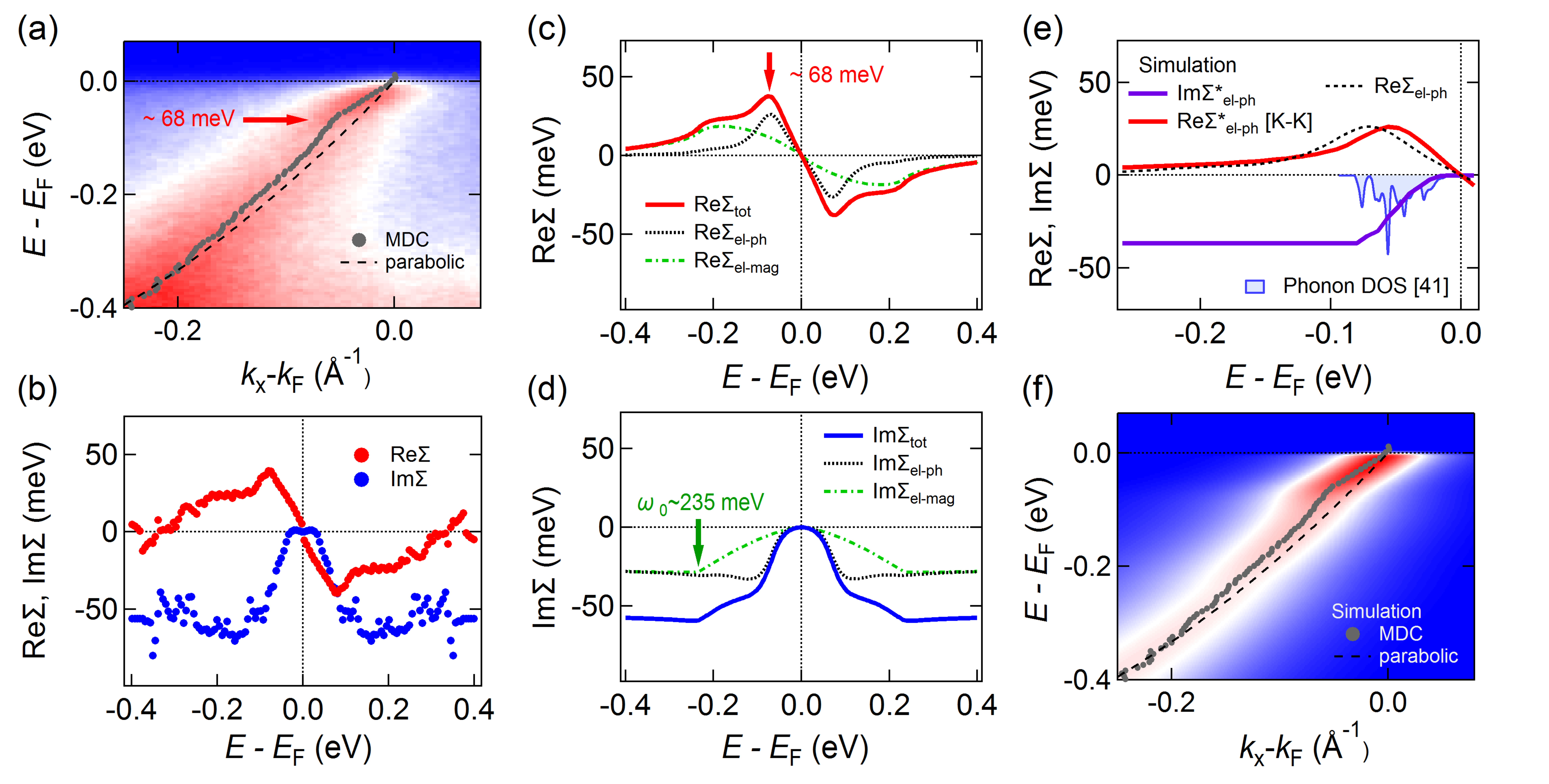}
\caption{(color online) (a) ARPES band dispersion along  $\Gamma^\prime$-X$^\prime$ direction in Fig. \ref{f2}(c). 
The grey solid line and the dashed line indicate the fitted results of  MDCs and the bare parabolic band dispersion without the electron-boson coupling. 
(b) Experimental real part of self-energy Re$\Sigma$ and imaginary part of self-energy Im$\Sigma$ deduced from the ARPES band dispersion of (a). 
Simulation of (c) Re$\Sigma$ and (d) Im$\Sigma$ based on the two-boson model with the electron-phonon (Green dashed-and-dotted line) and electron-magnon interaction (Black dashed line). 
(e) Calculated self-energy using the previously reported phonon-DOS \cite{Kim2012}. 
The dotted line indicates the real part of self-energy Re$\Sigma\subm{\textrm{el-ph}}$ in (c) for comparison with simulation. 
(f) Simulated ARPES data using the self-energies in (c) and (d). 
} 
\label{f3}
\end{figure*}

To address more details of the QP state near $E\subm{F}$, we have deduced the ARPES band dispersion from the Lorentz fitting of the momentum distribution curves (MDCs). 
Figure \ref{f3}(a) shows the MDC peak position of the ARPES intensity along $\Gamma^\prime$-X$^\prime$ direction and exhibits the distinct kink structure around 68 meV, suggesting the effect of some electron-boson interactions. 
Here, we assumed the bare band dispersion, not including the electron-boson interaction, by the parabolic interpolation between the Fermi wavenumber ($k\subm{F}$) and the MDC peak positions ranging from -0.32 eV to -0.44 eV. 
The energy scale of the kink structure is very close to both the Debye temperature $\Theta\subm{D} = 750$ K (64.6 meV) obtained from the temperature dependence of the electrical resistivity and the $A\subm{1g}$ Raman mode of 587 cm$^{-1}$ (72 meV)  \cite{Lewis1997, Iliev1999}. 
Thus, a plausible candidate for the kink structure is the electron-phonon interaction. 

The Fermi velocity was estimated to be $v\subm{F} = 1.9 \times 10^5$ m/s, which is within the range of previously reported values from 0.82 $\times 10^5$ m/s to 3.8 $\times 10^5$ m/s \cite{Zou2008, Bisti2017, Bheemarasetty2025}. 
Furthermore, the obtained $v\subm{F}$ is smaller than the reported value of $v\subm{F} = (3.8\pm0.2)\times 10^5$ m/s along $\Gamma$-X direction in the SX-ARPES \cite{Bisti2017}, suggesting that the improvement of the energy resolution due to the lower-energy incident light enabled its observation for the renormalized effect. 

The ARPES intensity contains not only the shape of the band dispersion deduced from the Lorentz fitting of MDCs, but also further information on electron-boson interactions. 
From the spectral width and the peak positions of MDCs, we have evaluated the real and imaginary parts of self-energy (Re$\Sigma$ and Im$\Sigma$) as shown in Fig. \ref{f3}(b). 
Here, Re$\Sigma$ was deduced from the difference between the experimental band dispersion and the bare band. 
Im$\Sigma$ was estimated from the Lorentz width $\delta k$ by following the relation Im$\Sigma = -\hbar v\subm{F}^0 \delta k/2$. 
The real part Re$\Sigma$ shows the peak structure around 68 meV, indicating the strong deviation from the bare band and the broad hump structure around 200 meV. 
These two features imply the two different bosonic modes coupled with electrons. 
On the other hand, the imaginary part Im$\Sigma$ is directly related to the QP lifetime, and the variation of Im$\Sigma$ corresponds to the change of the QP lifetime. 
Reflecting the feature of Re$\Sigma$, the imaginary part Im$\Sigma$ exhibits the drastic reduction around 58 meV, although there is no appreciable change of  Im$\Sigma$ around 200 meV.

In order to clarify the origin of the structure of the experimentally obtained self-energies, we have simulated the self-energy on the basis of the model including the electron-phonon and electron-magnon interactions characteristic of the half-metallic ferromagnet CrO$_2$.
The imaginary part Im$\Sigma$ is given by
\begin{align*}
{\rm{Im}}\Sigma(\omega) = -\pi \int^{\infty}_0 \alpha^2 F(\omega') [2n(\omega') + f(\omega'+\omega) \\+ f(\omega'-\omega)]d\omega' 
\end{align*}
by considering the electron-boson interaction, where $n(\omega)$ and $f(\omega)$ are the Bose-Einstein and Fermi-Dirac distribution functions, respectively. 
Here, $n(\omega)$ is ignored due to the data collection at sufficiently low temperature $T$ = 18 K. 
To simulate the electron-magnon interaction $\Sigma\subm{\textrm{el-mag}}$, we have assumed the magnon density of states (DOS): $\alpha^2 F(\omega) \propto \omega^{1/2}$ with a cutoff at maximum energy $\omega_0$ \cite{Schafer2004, Hayashi2013, Mazzola2022, Rost2024}. 
Here, the cutoff energy is set to be $\omega_0 \sim 235$ meV. 
The imaginary part Im$\Sigma\subm{\textrm{el-mag}}$ follows $\omega^{3/2}$ for $\omega < \omega_0$ and becomes constant above $\omega_c$ as shown in the green dashed-and-dotted lines in Fig. \ref{f3}(d). 
The obtained Im$\Sigma\subm{\textrm{el-mag}}$ leads to the broad hump structure in Re$\Sigma\subm{\textrm{el-mag}}$ of Fig. \ref{f3}(c) through the Kramers-Kronig (K-K) transformation. 
In addition to the magnon contribution, we have set the other $\Sigma\subm{\textrm{el-ph}}$ with the peak feature around 68 meV in the real part of the self-energy and obtained Im$\Sigma\subm{\textrm{el-ph}}$. 
By using two different contributions of the bosonic modes, the total self-energy $\Sigma\subm{tot}$ reproduces the shapes of the experimentally obtained self-energy $\Sigma$ of Fig. \ref{f3}(b). 
Although the lineshape of Im$\Sigma\subm{tot}$ is mostly consistent with the experimentally obtained Im$\Sigma$, the shoulder structure around 180 meV in Im$\Sigma\subm{tot}$ could not be identified in Im$\Sigma$. 
This is because, in addition to the $d_{yz+xz}$ band being analyzed, the contribution from the $d_{yz-xz}$ band is included in the intensity around -0.3 eV to -0.4 eV. 
Actually, the broad intensity independent on the wave number can be seen around -0.4 eV in Fig. \ref{f3}(a).

The experimentally obtained self-energies of Fig. \ref{f3}(b) are successfully reproduced with the electron-magnon contribution and the other electron-phonon-like contribution, while the latter is phenomenologically given. 
To confirm the validity of this procedure, we have simulated the Re$\Sigma\subm{\textrm{el-ph}}$ using the calculated phonon-DOS as $F(\omega)$ \cite{Kim2012}. 
Figure \ref{f3}(e) shows the simulated results for Re$\Sigma\subm{\textrm{el-ph}}^*$ and Im$\Sigma\subm{\textrm{el-ph}}^*$ compared with Re$\Sigma\subm{\textrm{el-ph}}$ used in Fig. \ref{f3}(c). 
The simulated imaginary part shows a gradual change from $E\subm{F}$ up to $– 80$ meV, reflecting the shape of the calculated phonon-DOS. 
By applying the K-K transformation to the imaginary part Im$\Sigma\subm{\textrm{el-ph}}^*$, the real part Re$\Sigma\subm{\textrm{el-ph}}^*$ was calculated. 
The resultant Re$\Sigma\subm{\textrm{el-ph}}^*$ (red solid line) exhibits the same shape as the phenomenologically introduced Re$\Sigma\subm{\textrm{el-ph}}$.  
The peak position of the calculated Re$\Sigma\subm{\textrm{el-ph}}^*$ is located around $-56$ meV, which is slightly lower than the phenomenological Re$\Sigma\subm{\textrm{el-ph}}$ (black dashed line). 
Here, we deduced the coupling constant $\lambda\subm{\textrm{el-ph}} \sim 0.6$ using the slope of Re$\Sigma\subm{\textrm{el-ph}}$: $\lambda = - \partial \rm{Re} \Sigma (\omega)/\partial  \omega$. 
The coupling constant $\lambda\subm{\textrm{el-ph}}$ would be comparable to $\lambda_{xx} = 0.8$ from the Bloch-Gr\"{u}neisen analysis of $\rho$ considering the electron-phonon interaction \cite{Lewis1997}. 
Moreover, the effective mass enhancement factor is estimated to be $1 + \lambda\subm{\textrm{el-ph}} \approx 1.6$, which is smaller than 2.5 - 4.34 estimated from the ratio of the experimental specific-heat coefficient to that from band calculations \cite{Lewis1997, Tsujioka1997}. 
This difference implies significant contributions not only from the electron-phonon interaction but also from the electron-electron interaction, that is, the renormalization due to the electron-electron correlation from the bare band mass. 
Based on the validity of these two bosonic modes as determined by the above analysis, the ARPES spectrum was reproduced in Fig. \ref{f3}(f) using the self-energies of Figs. \ref{f3}(c) and \ref{f3}(d). 
Here, we applied an offset of Im$\Sigma$ to take into account the finite lifetime width of the observed spectrum. 
The simulated ARPES spectra show good agreement with the experimentally obtained  ARPES data of Fig. \ref{f3}(a) by assuming the electron-phonon and electron-magnon interactions.

\section{CONCLUSION}
In summary, we have investigated the electronic structure of the half-metallic ferromagnet CrO$_2$ by high-resolution ARPES. 
Although the observed FSs and band dispersions are consistent with the previous reports, the distinct kink structure around 68 meV was observed in the ARPES band dispersion along the $\Gamma^\prime$–X$^\prime$ direction. 
The characteristic energy scale is comparable to the Debye temperature and the $A\subm{1g}$ phonon mode, suggesting the strong electron-phonon coupling. 
Furthermore, our self-energy analysis revealed characteristic features that can be well reproduced by considering both electron-phonon and electron-magnon interactions. 
These results demonstrate the renormalized QP dynamics in CrO$_2$ and offer new insights into the fundamental many-body interactions governing half-metallic ferromagnets.

\section*{ACKNOWLEDGMENTS}
The authors would like to thank T. Wakita for fruitful discussions. 
This work was supported by Japan Society for the Promotion of Science (JSPS) Grants-in-Aid for Scientific Research (KAKENHI Nos. 23K25822, 24K01161, and 25K07184). 
The synchrotron radiation experiment was performed with the approvals of Photon Factory (Proposal No. 2016G158 and No. 2018G127) and HSRC (Proposal No. 21BG032). 
H.F. acknowledges support from the JSPS Research Fellowship for Young Scientists. 
D.O and H.F are co-first authors and contributed equally to this work.

\section*{DATA AVAILABILITY}
The data that support the findings of this study are available from the corresponding author upon reasonable request.

\end{document}